\documentclass[%
 prd,twocolumn,nofootinbib,superscriptaddress
]{revtex4-2}
\usepackage{preamble}
\usepackage{bm}
\usepackage{subfigure}
 \usepackage{lineno}

\begin{document}
\title{Study of Galactic interactions and propagation properties of ultrahigh energy cosmic rays using the precision spectrum from the Pierre Auger Observatory}

\author{WeiKang Gao}
\email{gaowk@ihep.ac.cn}
\affiliation{Key Laboratory of Particle Astrophysics, Institute of High Energy Physics, Chinese Academy of Sciences, 100049 Beijing, China}
\affiliation{University of Chinese Academy of Sciences, 100049 Beijing, China}
\affiliation{Tianfu Cosmic Ray Research Center, 610000 Chengdu, Sichuan,  China}

\author{Dan Li}%
\email{lidan@ihep.ac.cn}
\affiliation{Key Laboratory of Particle Astrophysics, Institute of High Energy Physics, Chinese Academy of Sciences, 100049 Beijing, China}
\affiliation{University of Chinese Academy of Sciences, 100049 Beijing, China}
\affiliation{Tianfu Cosmic Ray Research Center, 610000 Chengdu, Sichuan,  China}

\author{Hua Yue}
\affiliation{Key Laboratory of Particle Astrophysics, Institute of High Energy Physics, Chinese Academy of Sciences, 100049 Beijing, China}
\affiliation{University of Chinese Academy of Sciences, 100049 Beijing, China}
\affiliation{Tianfu Cosmic Ray Research Center, 610000 Chengdu, Sichuan,  China}

\author{Wei Liu}
\affiliation{Key Laboratory of Particle Astrophysics, Institute of High Energy Physics, Chinese Academy of Sciences, 100049 Beijing, China}
\affiliation{Tianfu Cosmic Ray Research Center, 610000 Chengdu, Sichuan,  China}

\author{HongBo Hu}
\affiliation{Key Laboratory of Particle Astrophysics, Institute of High Energy Physics, Chinese Academy of Sciences, 100049 Beijing, China}
\affiliation{University of Chinese Academy of Sciences, 100049 Beijing, China}
\affiliation{Tianfu Cosmic Ray Research Center, 610000 Chengdu, Sichuan,  China}

\date{\today}

\begin{abstract}
The AUGER Collaboration has recently published the precise energy spectrum of cosmic rays above 1 EeV, which exhibits rich and interesting features. These features provide an opportunity to investigate the galactic propagation and interaction of ultra-high-energy cosmic rays (UHECRs). 
While the classic dip model,  which describes the UHECR propagation and interaction with CMB in extragalactic space by theoretical calculation, well explains the cutoff behaviour of the highest high energy spectrum of AUGER, it fails to describe the low energy features. It indicates that the galactic propagation and interaction of UHECR might play an important role in modulating the low energy side of AUGER spectrum. With the disk-halo diffusion model,  
this work demonstrates excellent agreement with the energy spectrum measured by AUGER and supports that the diffusion coefficient follows a power-law with an index of $1.81^{+0.05}_{-0.04}$ in the galaxy at least above $5\times10^{18}$eV. The inferred diffusion coefficient is one to two order of magnitude smaller than these extrapolated from low energy studies when conventional model parameters are adopted. Study of the modulation effect at cluster of galaxy level might be needed.
\end{abstract}

\maketitle

\section{Introduction}\label{sec1}

Significant progress has been made in the study of galactic cosmic ray propagation. The CREAM-I~\citep{ahn2008measurements}, TRACER~\citep{obermeier2011energy}, PAMELA~\citep{adriani2014measurement}, AMS-02~\citep{PhysRevD.91.063508,AMS:2015tnn,AMS:2017seo} NUCLEON~\citep{grebenyuk2019secondary} and DAMPE~\citep{dampe2022detection} experiment have promoted the development of diffusion model\citep{strong2007cosmic,Evoli_2017,vladimirov2011galprop}
of cosmic rays by providing precise measurements of the cosmic ray energy spectrum and the B/C ratio in the GeV-TeV range. Several experiments~\citep{bartoli2015argo,amenomori2005large,aartsen2016anisotropy} have also indicated that the anisotropy amplitude of cosmic rays is lower by an order of magnitude than what is predicted by the diffusion model. Additionally, the observation of TeV-Halo by the HAWC experiment~\citep{HAWC:2017kbo} indicates there is a slow diffusion region near the pulsar, and another analysis\citep{Bao_2019} suggests that slow diffusion regions might be common around pulsars.

Interesting discoveries have also been revealed at higher energy. Observations by experiment such as AUGER~\citep{PierreAuger:2010gfm,kampert2012highlights}, TA~\citep{TelescopeArray:2012qqu,TelescopeArray:2018xyi} or HiRes~\citep{SOKOLSKY200967} have extended the detection of Ultra-High-Energy Cosmic Rays (UHECRs) to 100EeV. In the analysis of UHECRs components~\citep{KAMPERT2012660}, experiments have identified a lighter “ankle” ($\sim$5EeV), implying that cosmic rays around the ankle may originate from extragalactic sources. This observation aligns with the anisotropic observations of UHECRs~\citep{PierreAuger:2014gko} that the dipole phase begins to reverse the pointing direction from galactic center to anti-galactic center when the energy exceeds EeV. Moreover, the recent precise energy spectrum measurement results~\citep{PierreAuger:2020kuy} from the AUGER collaboration have unveiled a rich and interesting structure in the cosmic ray spectrum. According to the latest observations of AUGER, the cosmic ray energy spectrum becomes harder at $5.0\times10^{18}$eV with the spectral index changing from -3.29 to -2.51.  At $1.3\times10^{19}$eV, the spectrum marks a softening as the spectral index transitions from -2.51 to -3.05, and the energy spectrum sharply drops to -5.1 at $4.6\times10^{19}$eV. The mass composition measurement of UHECRs by AUGER experiment can be found in \citep{PierreAuger:2023ebl} that protons predominate around the ankle, while TA\citep{Jui2011CosmicRI} and Hires\citep{PhysRevLett.104.161101} support that UHECRs are almost entirely composed of protons.

Presently, there are two mainstream models that demonstrate the spectral features above the ankle. Based on the observation that the average mass number of UHECRs becomes heavier as the energy increases, the mixed-composition model was proposed that the acceleration of cosmic rays is Z-dependent~\citep{2008ICRC....4..253A,2014JCAP...10..020A,aab2017combined}. Different components have different cutoff energies, and the sum of each nucleus forms the observed energy spectrum. Although this model can explain the energy spectrum and composition spectrum measured by AUGER, it requires a very hard spectral index of $E^{-1}\sim E^{-1.6}$ ~\citep{mollerach2019ultrahigh} at the source.  Another model, known as the "dip" model~\citep{1988A&A...199....1B,BEREZINSKY2014120, 2005PhLB..612..147B,2006PhRvD..74d3005B}, focuses primarily on protons.
Due to the interaction between extragalactic ultra-high-energy protons and CMB photons, the production of electron-positron pairs and pions with higher energy thresholds occurs. The energy loss resulting from these interactions accurately explains the energy spectrum above 10 EeV~\citep{harari2014anisotropies}. However, this model fails to satisfactorily account for the features of spectrum.

In this work, energy dependent galactic diffusion of UHECRs is studied under the dip model.  As lower energy cosmic rays penetrate galactic halo less efficiently than those with higher energies, the spectrum is expected to be suppressed at lower energy which may cure the opposite behavior of dip model.  The paper is organized in following way, section \ref{sec2} introduces the dip model plus the description of galactic propagation. Section \ref{sec3} presents the fit of spectrum with the data of AUGER and section \ref{sec:Conclusion and Discussion} concludes the paper and provides a discussion of the findings.

\begin{figure}
    \centering
    \includegraphics[width=0.4\textwidth,height=6.4cm]{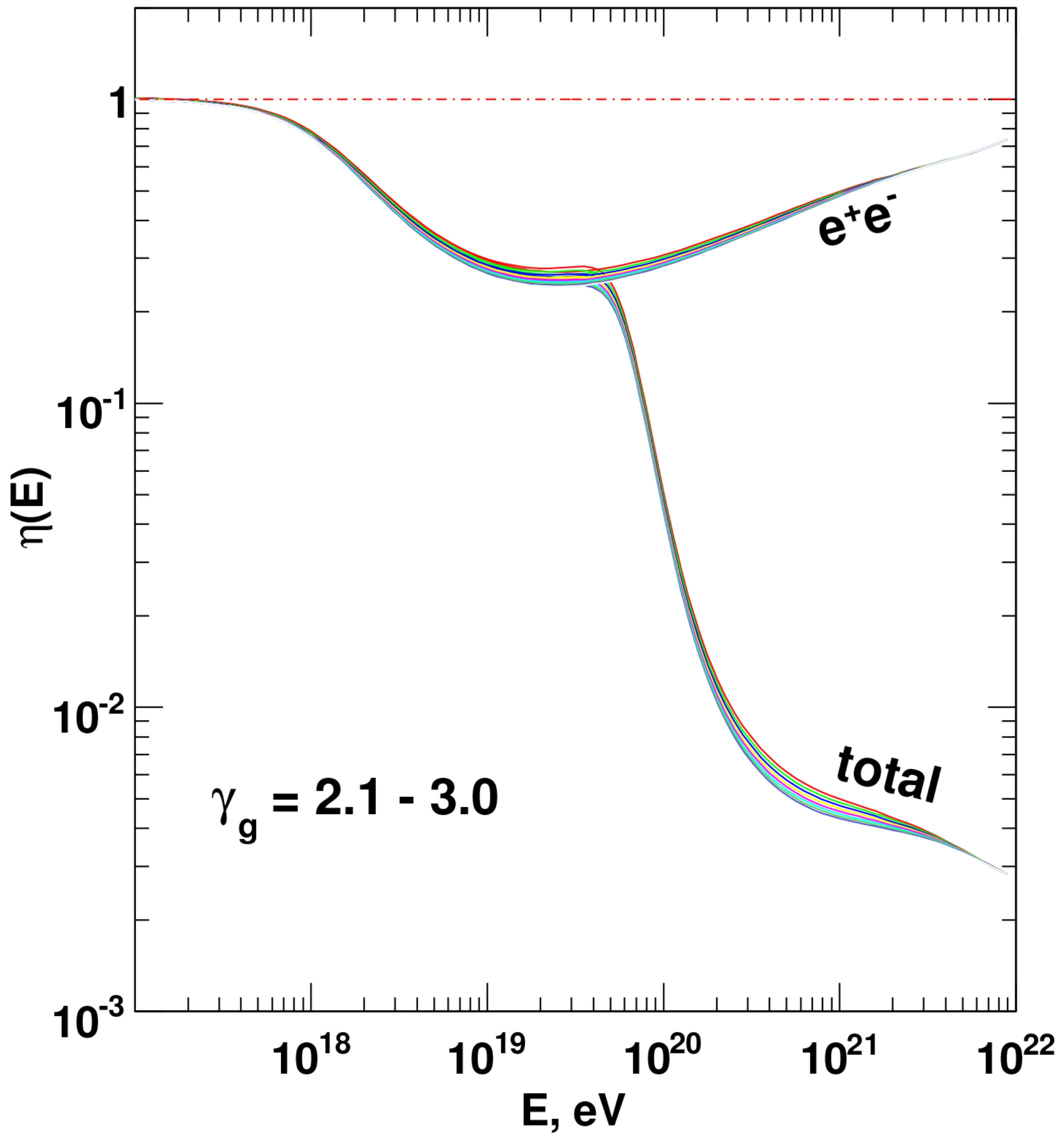}
    \caption{ Modification factors $\eta(E)$ for different generation index from ~\citep{ALOISIO2012129} (Reproduced with permission, courtesy of Elsevier.). The red dotted line shows the $\eta(E)$ for the adiabatic propagation of extragalactic cosmic rays, and the solid line shows the $\eta(E)$ under the energy loss with only the $e^{+}e^{-}$ pair production, or both $e^{+}e^{-}$ and $\pi^{+}$ production, where different colors represent different spectrum index of the extragalactic sources from -2.1 to -3.0. }
    \label{fig:enter-label}
\end{figure}

\section{Analysis method}
\label{sec2} 
The dip model~\citep{1988A&A...199....1B, BEREZINSKY2014120, 2005PhLB..612..147B, 2006PhRvD..74d3005B} describes the interaction and energy loss of the protons on their way from sources to our galaxy. Assuming that the source of UHECRs is far away from the galaxy and the entire universe is transparent, the UHECRs will only experience the redshift effect and energy loss, which is due to the interaction
\begin{equation}
    p + \gamma_{CMB}\longrightarrow \pi^{+} + X
\end{equation}
and 
\begin{equation}
     p + \gamma_{CMB}\longrightarrow e^{+}+e^{-} + p
\end{equation}
These process can be accurately calculated by particle physics and result in a significant cut-off of the proton flux at about 50EeV\citep{ALOISIO2012129}, namely the GZK cutoff. 

For other nuclei, the main energy loss process is
\begin{equation}
    A+ \gamma_{CMB} \longrightarrow (A-nN) + nN
\end{equation}
These processes have limited effects on the structure of energy spectrum and occur at lower energies\citep{ALOISIO201373}, so the dip model mainly focuses on the propagation of protons and assume that protons dominate in ultra high energies.

Ref~\citep{1988A&A...199....1B,ALOISIO2012129} calculated the modification factor $\eta (E) = \psi_{L}(E)/\psi_{s}(E)$ between observed local intensity near the Milky Way $ \psi_{L}(E)$ and averaged intensity from source $\psi_{s}(E)$.

The factor $\eta (E)$ is found to be approximately insensitive to the source spectrum index. Figure \ref{fig:enter-label} is taken from ~\citep{ALOISIO2012129} and shows the modification factor with different source spectral index from -2.1 to -3.0. The red dotted line shows the $\eta(E)$ for the adiabatic propagation of extragalactic cosmic rays, and the solid line shows the $\eta(E)$ under the energy loss with only the electron pair production, or both $e^{+}e^{-}$ and $\pi^{+}$ production.

 The galactic diffusion of the UHECRs is a possible solution, which is ignored in the dip model.
 According to the modified weighted slab model in quasi-linear diffusion theories ~\citep{2018JCAP...07..051G,2019JPhCS1181a2039K}, the energy dependence of the diffusion coefficient generally follows a double power-law. At lower energy where resonant scattering applies, the power law index is typically 1/3 when turbulence has a Kolmogorov spectrum\citep{kolmogorov1941local}, or 1/2 for a Krachin spectrum\citep{kraichnan1965inertial}. At higher energy where quasi-ballistic diffusion applies, the power law index switches to 2. In any case, the observation of cosmic rays will be significantly implicated. 

To study this process, we employ a 1-D space model known as the disk-halo diffusion model\citep{Jones_2001}, or called weighted slab model\citep{Jones_2001,Roberto_Aloisio_2013}. The model assumes a large radius for the Milky Way, allowing us to only focus on the z-direction. The galaxy disk is filled with interstellar gas with a half-thickness parameter of $h$. The halo, characterized by a half-thickness of $L$, is assumed to have a same diffusion coefficient as of the disk. Within the disk, cosmic rays interact with the galactic gas, while in the halo, cosmic rays diffuse isotropically. out of the halo, the intensity of cosmic rays is set to $\psi_{L}(E)$ according to the dip model. Consequently, the propagation equation in cylindrical coordinates is
\begin{equation}
\frac{\partial \psi}{\partial t}  =  \frac{\partial }{\partial z} (K(E)\frac{\partial\psi }{\partial z} )-2h\Gamma \delta (z)\psi + Q(z,p) 
- \frac{1}{p^{2}}\frac{\partial}{\partial p}[p^{2}(\frac{dp}{dt})\psi]
\end{equation}
where $\psi$ is the intensity of cosmic rays, $K(E)$ is defined as the diffusion coefficient for the halo, and $\Gamma = \beta cn\sigma_{int}$ is the rate of interaction between cosmic rays and gases on the disk. The $Q(z,p)$ is the injection rate of sources and the $\frac{dp}{dt}$ is the energy loss rate of cosmic rays. Because the energy of UHECRs is very high, the energy loss in the galaxy can be negligible. On the other hand, the UHECRs originate from extra galaxy, leaving no source term in the equation. The equation can be simplified
\begin{equation}
 \frac{\partial \psi}{\partial t}  =  \frac{\partial }{\partial z} (K(E)\frac{\partial\psi   }{\partial z} )-2h\Gamma \delta (z)\psi 
\end{equation}
 According to the proton-proton inelastic cross section $\sigma_{int}$ in DRAGON2\citep{Evoli_2018}, $\Gamma$ is regarded to follow a power-law with index $\sim 0.05$, so we set $\Gamma = \Gamma_{0} E^{0.05}$. 
In the case of steady state, $\frac{\partial\psi}{\partial t} = 0$, general form of the solution is denoted as $\psi =Az+B$. At $z=0$ and $z=L$, the flux continuity equations at boundary are set to be:
\begin{equation}
 \begin{cases}
 \lim_{\xi  \to 0} \int_{-\xi }^{\xi } \frac{\partial }{\partial z} (K(E)\frac{\partial\psi   }{\partial z} )-2h\Gamma \delta (z)\psi dz =0 & \text{ if } z=0 \\
    A L+B=\psi_{L}(E) & \text{ if } z=L \\

\end{cases}
\end{equation}

Finally, the intensity of cosmic rays on the earth is 

\begin{equation}
    \psi (z=0,E)=\frac{\psi_{L}(E) }{\frac{hL\Gamma }{K(E)}+1 } 
\end{equation}
where $\psi _{L}$is the injection spectrum from extra galaxy given by the dip model. If we focus on the effect of the Galaxy, $\eta_{G}  = \psi (z=0,E)/\psi _{L}(E)  $
\begin{equation}
    \eta_{G}(E) =\frac{1}{\frac{hL\Gamma }{K(E)}+1}\label{equ6}
\end{equation}
$\eta_{G}$ can be defined as a factor of galactic modulation. The diffusion coefficient $K$ is considered as a double power-law 
\begin{equation}
    K(E) = K_{0}(\frac{E}{5EeV})^{\delta_l}(1+\frac{E}{E_s})^{\delta-\delta_l}
\end{equation}
 $\delta_l$ represents the power index of the diffusion coefficient for low-energy, while $E_s$ is the energy at which the index changes. If we assume that the transmission energy $E_s$ is lower than the energy range we are concerned about, the $K$ can be described by a single power law $K(E) = K_{0}(\frac{E}{5EeV})^{\delta}$. 
Because of the significant degeneracy among $\Gamma_{0}$, $K_{0}$, $h$, and $L$, it is necessary to fit $\Gamma_{0}hL/K_{0}$ as a unified parameter.

\section{Results}
\label{sec3}
\begin{figure*}
    \centering
    \subfigure[]{\includegraphics[width=0.46\linewidth]{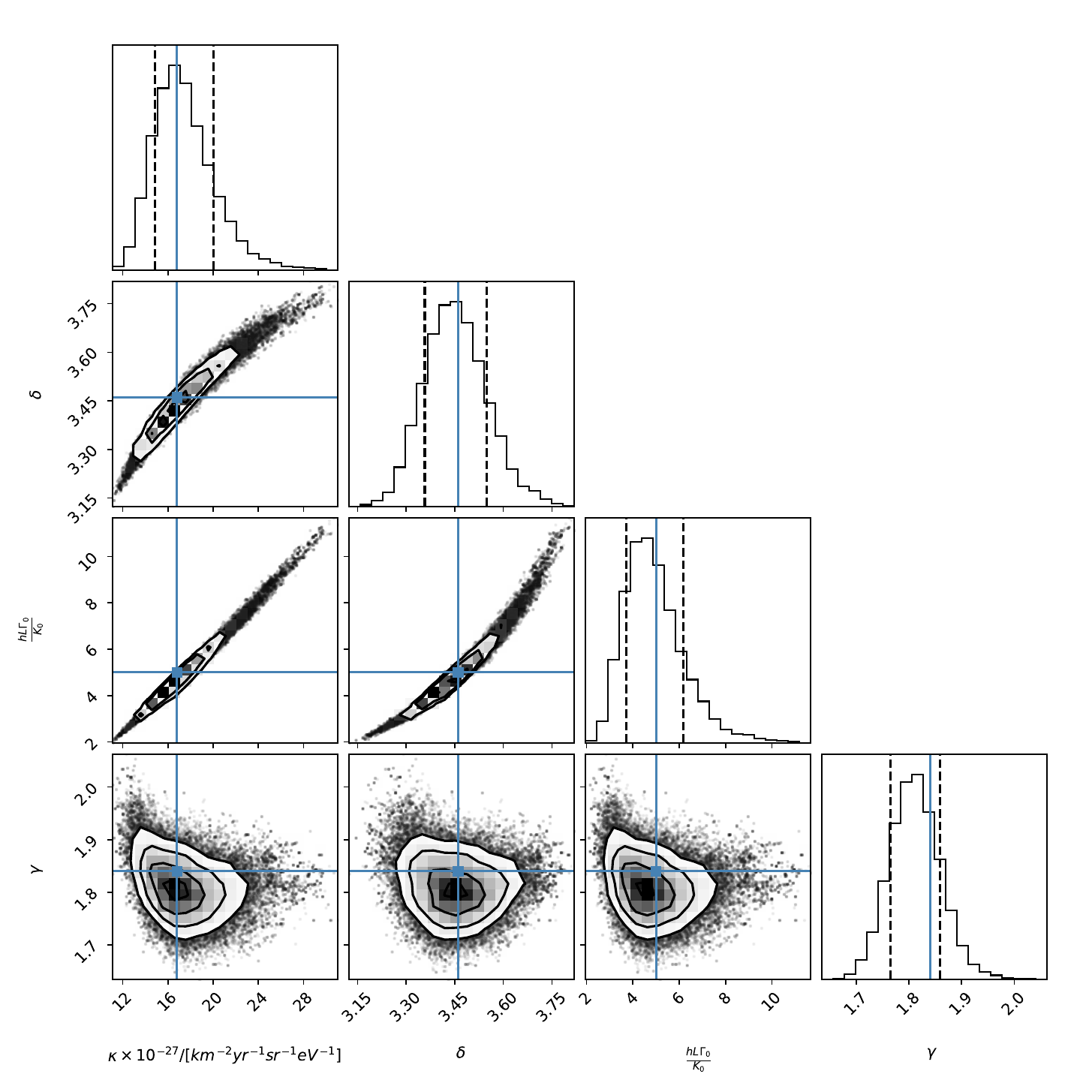}}
    \subfigure[]{\includegraphics[width=0.45\linewidth]{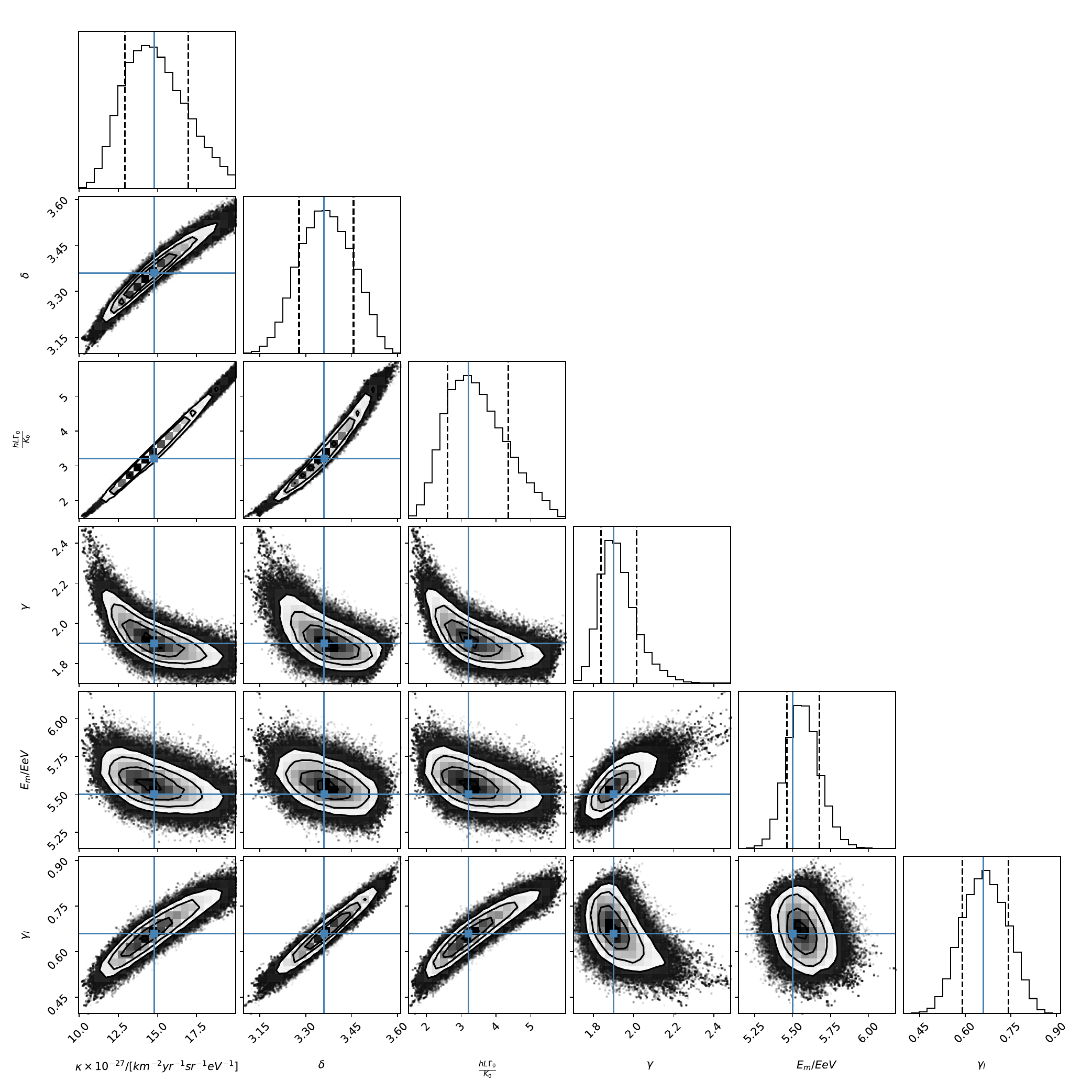}}
    \caption{Two-dimensional correlation distributions of parameters, and different contours represent different confidence intervals, namely 1-$\sigma$, 2-$\sigma$ and 3-$\sigma$. The blue line in the figure indicates the starting position of the parameters. (a) shows the distribution in LTE scenario, and figure (b) shows the distribution in HTE scenario.}\label{fit}
\end{figure*}

\begin{figure*}
    \centering
    \subfigure[]{\includegraphics[width=0.43\textwidth]{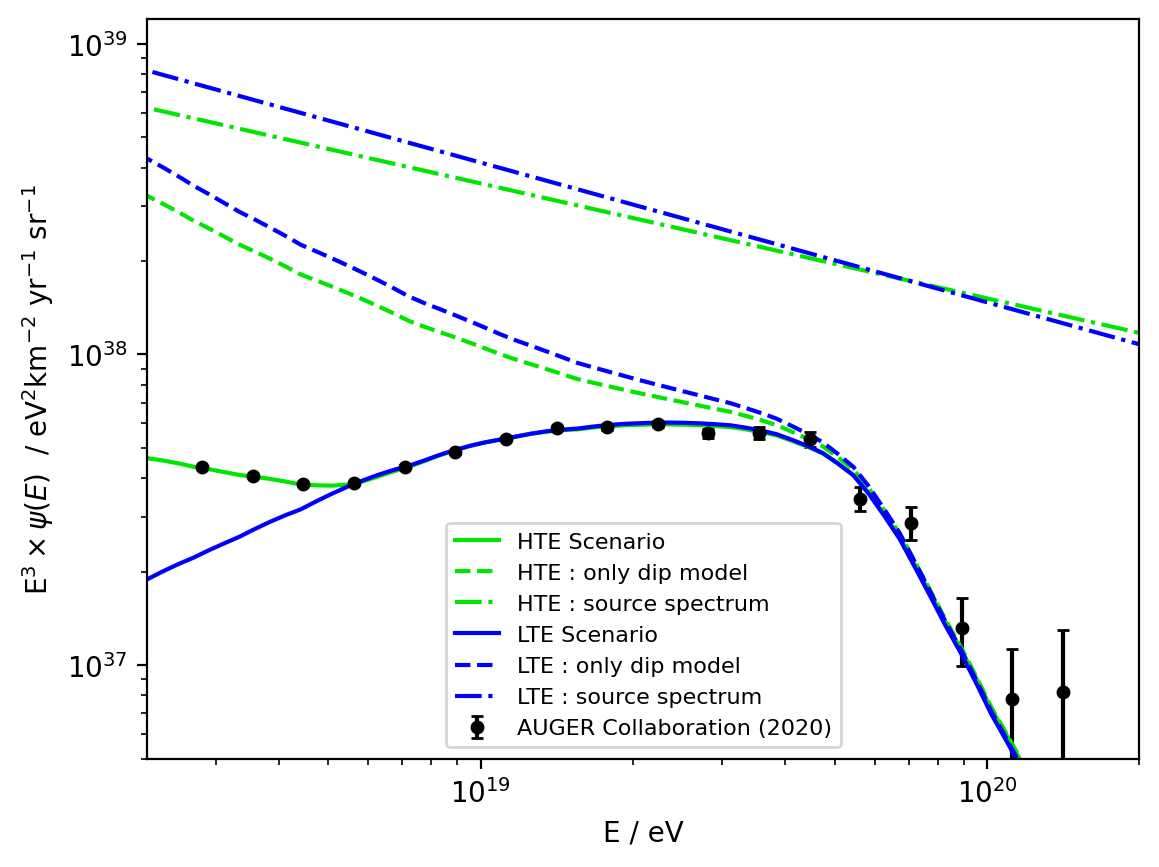}}
    \subfigure[]{\includegraphics[width=0.50\textwidth]{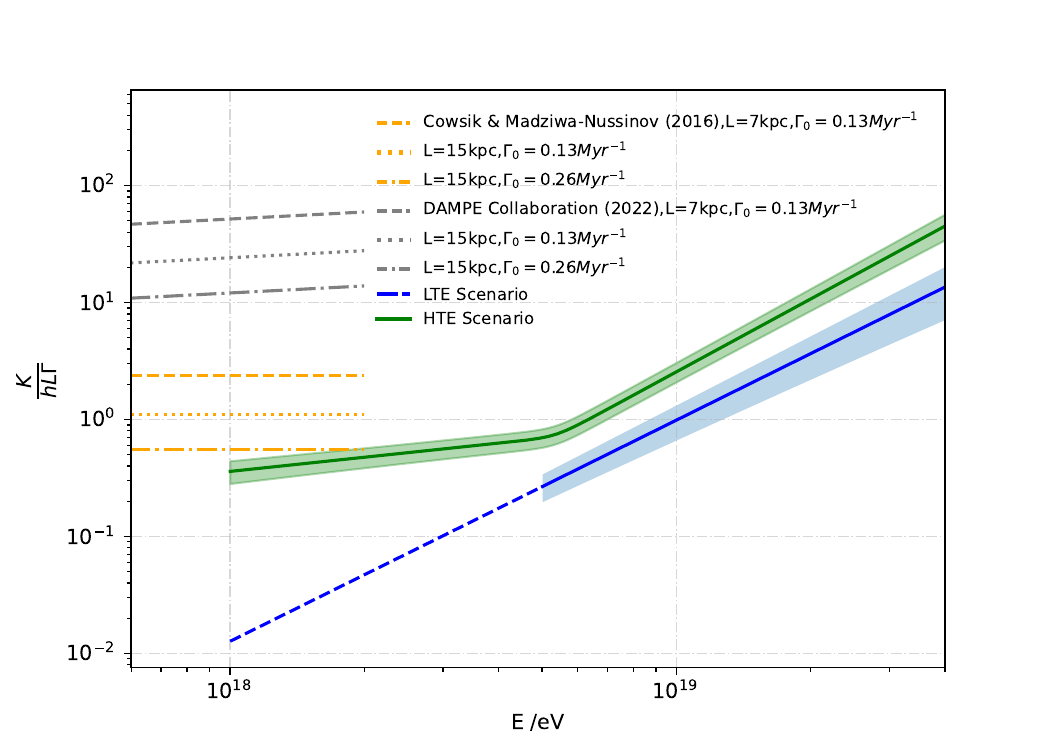}}
    \caption{(a) Fitted spectrum of different scenarios comparing to the measurement from AUGER. The scale of y-axis is multiplied with $E^{3}$. Green line shows that extragalactic cosmic rays dominate above 1 EeV, and blue line shows that extragalactic cosmic rays dominate above 5 EeV. The corresponding dot-dash lines and dash lines refer to the source spectrum and the spectrum with only the dip model in fitting parameters, respectively. (b) The spectrum of $K(E)/hL\Gamma$, blue line and green line represent the LTE and HTE scenario, respectively. We also present the extrapolation spectra of $K(E)/hL\Gamma$  predicted by two low-energy models, with the orange line from \cite{cowsik2016spectral} and the gray line from \citep{dampe2022detection}. The parameters for these two lines are fixed at $L = 7 kpc$ or $L=15kpc$, $\Gamma_0 =0.13Myr^{-1}$ or $\Gamma_0 =0.26Myr^{-1}$  and $h = 0.15kpc$.}

    \label{fig:1}
\end{figure*}

\begin{table*}
     \renewcommand{\arraystretch}{1.5}
    \centering
    \caption{ Fitting results of parameters and $\chi^{2}$ of fitting in the model. }
    \begin{ruledtabular}
    \begin{tabular}{cccccccc}\label{tab:1}
       & $\frac{hL\Gamma_{0}}{K_{0}} $ & $\delta$ &  $\kappa/[km^{-2}yr^{-1}sr^{-1}eV^{-1}] $ & $\gamma$ & $\delta_l$&$E_s/[eV]$&$\chi^{2}/d.o.f.$\\
    \colrule
    LTE& $4.88^{+1.50}_{-1.02}$ & $1.81^{+0.05}_{-0.04}$ & $1.74^{+0.31}_{-0.22}\times10^{-26}$ &  $3.45^{+0.10}_{-0.09}$ & ... & ... & $16.3/(14-4)$ \\
    HTE& $3.45^{+0.96}_{-0.78}$ &$1.91^{+0.10}_{-0.07} $ & $1.49^{+0.22}_{-0.19}\times10^{-26}$ & $3.37^{+0.09}_{-0.09}$ & $0.67^{+0.07}_{-0.07}$ & $5.56^{+0.11}_{-0.11}\times10^{18}$ & $40.0/(17-6)$ \\
\end{tabular}
\end{ruledtabular}
\end{table*}

With the galactic diffusion model, we have the $\psi(E) = \eta_{G}(E) \eta(E) \psi_{s}(E)$ to fit the spectrum of AUGER. The spectrum of extragalactic source has not been determined and is set to be a power-low $\psi_{s}(E)=\kappa E^{-\gamma}$. Therefore, this model has a total of 6 parameters: $\Gamma_{0}hL/K_{0}$, $\delta$, $\delta_l$, $E_s$, $\kappa$ and $\gamma$.

As the upper limit of energy attainable by galactic cosmic rays remains unknown, we subsequently discuss two scenarios. In low transmission energy(LTE) scenario, galactic cosmic rays can be accelerated to ultra-high energies, and the contribution of extragalactic cosmic rays starts to dominate from the “ankle”. In this case, we fit the model parameters only with energy spectrum of AUGER above 5EeV. 

 The second scenario is the extreme scenario. Extragalactic cosmic rays begin to dominate at lower energies, and the entire energy spectrum from 1EeV to 100EeV is contributed by extragalactic cosmic rays. This scenario is quite different from the theoretical prediction and is discussed here as a limit case to ensure the robustness of our conclusions.

We use the maximum-likelihood method to fit these parameters. The current intensity near the Earth calculated by the model is $F^{th}(E) = c/4\pi \times \psi(0,E)$, while the flux measured by AUGER experiment is $F^{exp}(E)$. The likelihood function is defined as
\begin{equation}
    \mathcal{L}(\Theta) = exp(-\frac{1}{2}\chi^{2}(\Theta))
\end{equation}
with the $\chi^{2}(\Theta)$ bulit from the data and model,
\begin{equation}
    \chi^{2}(\Theta) = \sum^{N^{data}}_{i=1}(\frac{F^{th}_{i}(\Theta)-F^{exp}_{i}}{\sigma_{i}})^{2}
\end{equation}
in which $F^{exp}_{i}$ and $\sigma_{i}$ are the measured value and standard deviation in experiment, and $F^{th}$ is the theoretical expectation under the certain parameter set $\Theta$. In the parameter estimation, the best-fit parameters are obtained using Markov chain Monte Carlo (MCMC) sampling methods, where at equilibrium the chain contains a set of samples pointing to the best-fit values in the parameter space. With this chain, the correlation distributions and $1-\sigma$ ranges can be derived as shown in figure \ref{fit}. The python package emcee\citep{Foreman-Mackey_2013} and corner are involved in the fitting process.

The parameter best-fit values and error ranges are displayed in Table \ref{tab:1}. In the LTE scenario, because the transmission energy is lower than the lowest energy of the spectrum, a single power law $K(E)$ is adopted. In the high transmission energy(HTE) scenario, the inflection of the diffusion coefficient generates an "ankle" on the energy spectrum, which means that the ankle corresponds to the transmission energy $E_s$.

The parameters show slight difference between the two scenarios. For the first parameter $hL\Gamma_{0}/K_{0}$, fitting results are $4.88^{+1.50}_{-1.02}$ in the LTE and $3.45^{+0.96}_{-0.78}$ in the HTE scenario.
For the spectrum index of $K(E)$, $\delta $ is fitted to be $1.81^{+0.05}_{-0.04}$ in LTE and $1.91^{+0.10}_{-0.07}$ in HTE, which is consistent with quasi-linear diffusion theory ~\citep{reichherzer2022regimes} in high energy. In this theory, UHECRs diffuse in the quasi-ballistic regime(QBR), where the diffusion coefficient is proportional to $E^{2}$, see appendix A. 
The parameters $\kappa$ and $\gamma$ describe the energy spectrum of extragalactic sources, where $\kappa$ is the normalization factor of the spectrum, which is around $10^{-26} km^{2}yr^{-1}sr^{-1}eV^{-1}$ in both scenarios. The parameter $\gamma$ represents the injection spectral index of extragalactic sources, with a value of 3.45 in the LTE scenario and 3.37 in the HTE scenario. 
In the HTE scenario, the origin of ankle is due to a break in the $K(E)$ spectrum. Therefore, the fitted value of the break energy $E_s$ for the diffusion coefficient is around the ankle, at $5.56\times 10^{18}$ eV. The spectral index $\delta_l$ of $K(E)$ at low energies is determined by the sharpness of the ankle, with a fitted value of 0.67. 
The blue line and green line in figure \ref{fig:1} (a) show the fitting result of this model. The small errors in the spectrum below the ankle contribute significantly to the $\chi^{2}$, resulting in a worse fit in the HTE scenario, while the LTE scenario can fit better.

At higher energies, the diffusion coefficient rapidly increases, allowing cosmic rays to escape the confines of galaxy which explains the softening of cosmic rays spectrum around 16EeV. When the energies go down, cosmic rays are more likely to be confined in the galaxy and interact with gases, resulting in a suppression of the cosmic rays spectrum. However, the variation of diffusion coefficient with energy also strongly affects the structure of energy spectrum as discussed in the HTE scenario. Figure \ref{fig:1}(b) shows the reciprocal of the fitted parameter $hL\Gamma_{0}/K_0(E)$ in both scenarios.

\section{Conclusion and Discussion}
\label{sec:Conclusion and Discussion}
Based on the dip model, the galactic diffusion effect to the UHECRs spectrum is studied. In this work, two scenarios of extragalactic contribution are discussed, and parameters in model have been fitted. With the best fitting spectral index $1.81^{+0.05}_{-0.04}$ and $1.91^{+0.10}_{-0.07}$ for diffusion coefficient, AUGER spectrum features can be well demonstrated. Relevant studies have predicted a power law change of $K$ at ultra high energies \citep{2018JCAP...07..051G,2019JPhCS1181a2039K,Reichherzer_2021}, which is also adopted in this model. In the extreme case, namely HTE scenario, the lower energy index of diffusion coefficient $\delta_l$ is fitted to be 0.67 and the break energy is at $5.56\times10^{18}eV$. Significantly, relevant works\citep{zank1998radial,wang2022turbulent} can be used to infer that the transmission energy $E_s$ should be under $EeV$, meaning that the LTE scenario is preferred by the quasi-linear diffusion theory. But it is correct to say that above 5 EeV, the power-law of the diffusion coefficient is around 2.

The values of these parameters are interesting. First, the height $L$ of the Galaxy halo and the thickness $h$ of the galaxy disk are not accurately measured, 7kpc and 0.15kpc here are representative values from the parameter ranges given in reference \citep{ackermann2012fermi,feng2016bayesian}. The thickness of the disk represents the concentration of matter in the Milky Way, and its size can be determined by the mass and gas density of the Milky Way. The height of the Milky Way refers to the region covered by the turbulent magnetic field of the Milky Way, and the cosmic rays in the region will undergo the diffusion process. The interstellar medium(ISM) surface density is taken as $h\times n \sim 0.15kpc\times 1cm^{-3}$, and the p-p inelastic cross section at 5EeV is taken as 140mb. The interaction ratio $\Gamma$ is calculated to be about $0.13Myr^{-1}$ at 5EeV, so the diffusion coefficients at 5EeV are $K_{5EeV} \sim 8.0\times10^{28} cm^{2}/s$ in LTE and $K_{5EeV}\sim2.2\times10^{29}cm^{2}/s$ in HTE scenario. Due to the degeneracy of the parameters in the model, $h$, $L$ and $\Gamma$ are proportional to the diffusion coefficient, for example, if UHECRs spreads in the range L=15kpc (i.e., Take the height of fermi bubble as the height of the Galaxy\citep{Yang:2022jck} ), the diffusion coefficient in the model will be twice as high as expected. The discussion of the different $L$ is shown in Figure\ref{fig:1} (b). On the other hand, LHAASO observed  diffuse gamma rays in the galaxy disk\citep{cao2023measurement} from 10 TeV to 1PeV, and the flux is three times as high as expected in the inner region and two times as high as expected in the outer region. The diffuse gamma rays on the galaxy disk are generally believed to come from cosmic rays interacting with the ISM, and the increased flux suggests that the interaction rate between cosmic rays and ISM may be greater. The dot-dash lines in Figure\ref{fig:1} (b) show the parameters with increased interaction rates $\Gamma_0 =0.26Myr^{-1}$. Because the uncertainty of the index of $\Gamma$ is very small compared to the $\delta$ and $\delta_l$, the power-law index is fixed at 0.05 and the uncertainty of interaction rate $\Gamma$ is ignored in this study.

It is very hard to make an accurate comparison between the diffusion coefficient obtained in this work and those from the observation of satellite experiments. The predicted  $K(E)$ is very small compared with the extended diffusion coefficient measured between GeV and TeV energies.
The DAMPE collaboration measured a power-law change in the B/C ratio at low energies ($\sim200GeV$), from 0.3 to 0.2. They proposed a break diffusion coefficient model, and the grey line in figure\ref{fig:1} (b) shows the extended diffusion coefficient with a power law of 0.2 at high energies. The model given in \citep{cowsik2016spectral} assumes that the diffusion coefficient of cosmic rays with energies $>200GeV$ is a constant, and the extended results are shown with orange line. 

 In this paper, the propagation in galaxy cluster is ignored, but there may still be modulation of UHECRs in the galaxy cluster. Cluster modulation has been investigated extensively, and it is believed to consist of diffusion and energy loss\citep{PhysRevD.71.083007,PhysRevD.104.063005}. 
 Interacting with CMB photons and intergalactic medium is mainly responsible for the energy loss of UHECRs, and the contributions of both are discussed in detail in the reference\citep{2023ApJ...957...80C}.
 On the other hand, turbulent magnetic fields in the galaxy cluster lead to the diffusion\citep{10.1111/j.1365-2966.2008.14132.x}. Such diffusion can trap the UHECRs for a longer time, which enhances the influence of the energy loss and allows them to interact sufficiently. If the cluster modulation exists, the modulation of the UHECR spectrum will be contributed by both the galaxy cluster and the galaxy, resulting in a larger diffusion coefficient in the galaxy. The cluster modulation of the UHECR would cascade down more very-high-energy gamma rays\citep{murase2012blazars}, which would harden the very-high-energy gamma ray spectrum of the cluster and could be observed in the future by experiments such as LHAASO (fig. 16 in Chap. 4 of ref \citep{bai2022large}).

The fitted spectral power-laws are -3.37 and -3.45, which are obviously soft compared to other theoretical cosmic ray sources, such as \citep{PhysRevD.99.103010,ALOISIO2012129}. These models suggest that the extragalactic injection spectrum should have a power-law between 2.1 and 2.7, which would explain the extragalactic source by shock acceleration. There is no reasonable explanation for the fitting results and have few experimental observations of extragalactic sources now. But the steeper energy spectrum corresponds to more extragalactic cosmic ray energy, which can be determined by further observations of extragalactic gamma ray.

Previous works such as \citep{10.1143/PTP.113.721} and \citep{osti_6709451} have also studied the galactic modulation for extragalactic cosmic rays. The dynamical wind model is adopted in both works. The first work proposes that the extragalactic cosmic rays contribute significantly to the formation of the knee region. The second work also illustrates the influence of galactic modulation on the extragalactic cosmic rays, and gives the analytical solution of cosmic ray density. They use different parameters in the dynamical wind model, so the modulation of extragalactic cosmic ray spectrum is different. The second work believes the suppression is close to the lower energy $\sim10GeV$.

We use similar methods, and think that the extragalactic cosmic rays should be cut off at low energies due to galactic modulation, except that they thought the cut-off occurs in the knee region, and we believe that the cut-off is above the ankle.

Finally, the effect of galaxy cluster on UHECR propagation is the key point in subsequent studies. The diffusion and interaction of UHECRs might be an effective solution for the spectral index problem in the mixed-composition model and requires to be considered in all kinds of extragalactic cosmic ray propagation study.

\begin{acknowledgements}
We thank XiaoJun Bi for valuable discussions about this work. We thank Astroparticle Physics for licensing the figure and the anonymous reviewers for their helpful comments. This work is supported by the Ministry of Science and Technology of China, National Key $R\&D$ Program of China (2018YFA0404203). This work is supported by National Natural Science Foundation of China (NSFC) (Nos. U2031110, 12333006).  

\end{acknowledgements}
\appendix
\section{Diffusion coefficient derivation in QBR regime}
Specific derivation of diffusion coefficient can be found in paper\citep{Subedi_2017}, only a brief derivation for the QBR regime is displayed subsequently.

The motion equation of particles in interstellar magnetic fields can be expressed using the Newton-Lorentz formula
\begin{equation}\label{equ:a1}
    \frac{d\mathbf{v}}{dt} = \frac{q}{\gamma mc}(\mathbf{v}\times (\mathbf{B}+\mathbf{b}))
\end{equation}
where the $\mathbf{v}$ is the velocity of particle, $q$ is the charge of particle, $\gamma$ is the Lorentz factor, and c is the velocity of light. The magnetic field is divided into two parts, where $\mathbf{B}$ refers to the constant background magnetic field, and $\mathbf{b}$ refers to the turbulent magnetic field.
Assuming that cosmic rays propagate isotropic in space, we can define spatial diffusion coefficient $K_{ii}$ and velocity space diffusion coefficient $D_{ii}$  by the TGK method \citep{taylor1922diffusion,Kubo1957}
\begin{equation}\label{equ:a2}
K_{ii} = \int_{0}^{\infty } dt<v_i(0)v_i(t)>  , i=x,y,z
\end{equation}
and
\begin{equation}\label{equ:a3}    
D_{ii} = \int_{0}^{\infty} dt<\frac{dv_i}{dt}(0)\frac{dv_i}{dt}(t)> , i=x,y,z
\end{equation}
They satisfy the equation\citep{Subedi_2017}
\begin{equation}\label{equ:a4}
    K_{ii} = \frac{v^{4}}{6D_{ii}}
\end{equation}
By substituting equation\ref{equ:a1} into equation\ref{equ:a3} and assuming the uncorrelation between particle velocity and magnetic field, the expression for $D_{ii}$ can be derived
\begin{equation}\label{equ:a5}
    D_{ii} = (\frac{q}{\gamma mc})^{2}\int_{0}^{\infty} dt \epsilon _{i\alpha\beta}\epsilon _{i\zeta\eta} <v_{\alpha}(0)v_{\zeta}(t)><B_{\beta}(0)B_{\eta}(\mathbf{x}(t))>
\end{equation}
When the energy of diffusing particles is very high(in QBR regime), the Larmor radius $r_{g}$ is significantly large, far exceeding the coherence length $l_{c}$ of the interstellar magnetic field ($r_{g}/l_{c} >> 1$). Therefore, the scattering angle of particles is very small, and the trajectory of particle is approximately a straight line. Consequently, the velocity correlation function $<v_{\alpha}(0)v_{\zeta}(t)> = v_{\alpha}v_{\zeta}$, and equation \ref{equ:a5} simplifies to
\begin{equation}\label{equ:a6}
    D_{ii} = (\frac{q}{\gamma mc})^{2}\epsilon _{i\alpha\beta}\epsilon _{i\zeta\eta}v_{\alpha}v_{\zeta}\int_{0}^{\infty} dt  <B_{\beta}(0)B_{\eta}(\mathbf{x}(t))>
\end{equation}
The coherence length of the magnetic field is defined as
\begin{equation}\label{equ:a7}
    l_{c} = \frac{c}{b^{2}}\int_{0}^{\infty} dt <B_{\beta}(0)B_{\eta}(\mathbf{x}(t))>
\end{equation}
Substituting equation\ref{equ:a7} into equation\ref{equ:a6}, and we have 
\begin{equation}\label{equ:a8}
    D_{ii} \propto \frac{q^{2}b^{2}l_c}{\gamma^{2}m^{2}} \propto E^{-2}
\end{equation}
Apply the equation \ref{equ:a8} in the equation \ref{equ:a4}
\begin{equation}
    K_{ii} \propto E^{2}
\end{equation}
which means that the spatial diffusion coefficient is proportional to the square of particle energy at very high energy. 

\bibliographystyle{unsrt}
\bibliography{ms}

\begin{thebibliography}{10}

\bibitem{ahn2008measurements}
HS~Ahn, PS~Allison, MG~Bagliesi, JJ~Beatty, G~Bigongiari, PJ~Boyle, TJ~Brandt, JT~Childers, NB~Conklin, Stephane Coutu, et~al.
\newblock Measurements of cosmic-ray secondary nuclei at high energies with the first flight of the cream balloon-borne experiment.
\newblock {\em Astroparticle Physics}, 30(3):133--141, 2008.

\bibitem{obermeier2011energy}
A~Obermeier, M~Ave, P~Boyle, Ch~H{\"o}ppner, J~H{\"o}randel, and D~M{\"u}ller.
\newblock Energy spectra of primary and secondary cosmic-ray nuclei measured with tracer.
\newblock {\em The Astrophysical Journal}, 742(1):14, 2011.

\bibitem{adriani2014measurement}
O~Adriani, GC~Barbarino, GA~Bazilevskaya, Roberto Bellotti, M~Boezio, EA~Bogomolov, M~Bongi, V~Bonvicini, S~Bottai, Alessandro Bruno, et~al.
\newblock Measurement of boron and carbon fluxes in cosmic rays with the pamela experiment.
\newblock {\em The Astrophysical Journal}, 791(2):93, 2014.

\bibitem{PhysRevD.91.063508}
Su-Jie Lin, Qiang Yuan, and Xiao-Jun Bi.
\newblock Quantitative study of the ams-02 electron/positron spectra: Implications for pulsars and dark matter properties.
\newblock {\em Phys. Rev. D}, 91:063508, Mar 2015.

\bibitem{AMS:2015tnn}
M.~Aguilar et~al.
\newblock {Precision Measurement of the Proton Flux in Primary Cosmic Rays from Rigidity 1 GV to 1.8 TV with the Alpha Magnetic Spectrometer on the International Space Station}.
\newblock {\em Phys. Rev. Lett.}, 114:171103, 2015.

\bibitem{AMS:2017seo}
M.~Aguilar et~al.
\newblock {Observation of the Identical Rigidity Dependence of He, C, and O Cosmic Rays at High Rigidities by the Alpha Magnetic Spectrometer on the International Space Station}.
\newblock {\em Phys. Rev. Lett.}, 119(25):251101, 2017.

\bibitem{grebenyuk2019secondary}
V~Grebenyuk, D~Karmanov, I~Kovalev, I~Kudryashov, A~Kurganov, A~Panov, D~Podorozhny, A~Tkachenko, L~Tkachev, A~Turundaevskiy, et~al.
\newblock Secondary cosmic rays in the nucleon space experiment.
\newblock {\em Advances in Space Research}, 64(12):2559--2563, 2019.

\bibitem{dampe2022detection}
Dampe Collaboration et~al.
\newblock Detection of spectral hardenings in cosmic-ray boron-to-carbon and boron-to-oxygen flux ratios with dampe.
\newblock {\em Science Bulletin}, 67(21):2162--2166, 2022.

\bibitem{strong2007cosmic}
Andrew~W Strong, Igor~V Moskalenko, and Vladimir~S Ptuskin.
\newblock Cosmic-ray propagation and interactions in the galaxy.
\newblock {\em Annu. Rev. Nucl. Part. Sci.}, 57:285--327, 2007.

\bibitem{Evoli_2017}
Carmelo Evoli, Daniele Gaggero, Andrea Vittino, Giuseppe~Di Bernardo, Mattia~Di Mauro, Arianna Ligorini, Piero Ullio, and Dario Grasso.
\newblock Cosmic-ray propagation with dragon2: I. numerical solver and astrophysical ingredients.
\newblock {\em Journal of Cosmology and Astroparticle Physics}, 2017(02):015, feb 2017.

\bibitem{vladimirov2011galprop}
Andrey~E Vladimirov, Seth~W Digel, Gudlaugur Johannesson, Peter~F Michelson, Igor~V Moskalenko, Patrick~L Nolan, Elena Orlando, Troy~A Porter, and Andrew~W Strong.
\newblock Galprop webrun: an internet-based service for calculating galactic cosmic ray propagation and associated photon emissions.
\newblock {\em Computer Physics Communications}, 182(5):1156--1161, 2011.

\bibitem{bartoli2015argo}
B~Bartoli, P~Bernardini, XJ~Bi, Z~Cao, S~Catalanotti, SZ~Chen, TL~Chen, SW~Cui, BZ~Dai, A~D’Amone, et~al.
\newblock Argo-ybj observation of the large-scale cosmic ray anisotropy during the solar minimum between cycles 23 and 24.
\newblock {\em The Astrophysical Journal}, 809(1):90, 2015.

\bibitem{amenomori2005large}
M~Amenomori, S~Ayabe, SW~Cui, LK~Ding, XH~Ding, CF~Feng, ZY~Feng, XY~Gao, QX~Geng, HW~Guo, et~al.
\newblock Large-scale sidereal anisotropy of galactic cosmic-ray intensity observed by the tibet air shower array.
\newblock {\em The Astrophysical Journal}, 626(1):L29, 2005.

\bibitem{aartsen2016anisotropy}
MG~Aartsen, K~Abraham, M~Ackermann, J~Adams, JA~Aguilar, M~Ahlers, M~Ahrens, D~Altmann, T~Anderson, I~Ansseau, et~al.
\newblock Anisotropy in cosmic-ray arrival directions in the southern hemisphere based on six years of data from the icecube detector.
\newblock {\em The Astrophysical Journal}, 826(2):220, 2016.

\bibitem{HAWC:2017kbo}
A.~U. Abeysekara et~al.
\newblock {Extended gamma-ray sources around pulsars constrain the origin of the positron flux at Earth}.
\newblock {\em Science}, 358(6365):911--914, 2017.

\bibitem{Bao_2019}
Yiwei Bao, Siming Liu, and Yang Chen.
\newblock On the gamma-ray nebula of vela pulsar. i. very slow diffusion of energetic electrons within the tev nebula.
\newblock {\em The Astrophysical Journal}, 877(1):54, may 2019.

\bibitem{PierreAuger:2010gfm}
J.~Abraham et~al.
\newblock {Measurement of the Energy Spectrum of Cosmic Rays above $10^{18}$ eV Using the Pierre Auger Observatory}.
\newblock {\em Phys. Lett. B}, 685:239--246, 2010.

\bibitem{kampert2012highlights}
Karl-Heinz Kampert.
\newblock Highlights from the pierre auger observatory, 2012.

\bibitem{TelescopeArray:2012qqu}
T.~Abu-Zayyad et~al.
\newblock {The Cosmic Ray Energy Spectrum Observed with the Surface Detector of the Telescope Array Experiment}.
\newblock {\em Astrophys. J. Lett.}, 768:L1, 2013.

\bibitem{TelescopeArray:2018xyi}
R.~U. Abbasi et~al.
\newblock {Depth of Ultra High Energy Cosmic Ray Induced Air Shower Maxima Measured by the Telescope Array Black Rock and Long Ridge FADC Fluorescence Detectors and Surface Array in Hybrid Mode}.
\newblock {\em Astrophys. J.}, 858(2):76, 2018.

\bibitem{SOKOLSKY200967}
Pierre Sokolsky.
\newblock Observation of the gzk cutoff by the hires experiment.
\newblock {\em Nuclear Physics B - Proceedings Supplements}, 196:67--73, 2009.
\newblock Proceedings of the XV International Symposium on Very High Energy Cosmic Ray Interactions (ISVHECRI 2008).

\bibitem{KAMPERT2012660}
Karl-Heinz Kampert and Michael Unger.
\newblock Measurements of the cosmic ray composition with air shower experiments.
\newblock {\em Astroparticle Physics}, 35(10):660--678, 2012.

\bibitem{PierreAuger:2014gko}
A.~Aab et~al.
\newblock {Depth of maximum of air-shower profiles at the Pierre Auger Observatory. II. Composition implications}.
\newblock {\em Phys. Rev. D}, 90(12):122006, 2014.

\bibitem{PierreAuger:2020kuy}
Alexander Aab et~al.
\newblock {Features of the Energy Spectrum of Cosmic Rays above 2.5\texttimes{}10$^{18}$ eV Using the Pierre Auger Observatory}.
\newblock {\em Phys. Rev. Lett.}, 125(12):121106, 2020.

\bibitem{PierreAuger:2023ebl}
Adila Abdul~Halim et~al.
\newblock {Highlights from the Pierre Auger Observatory}.
\newblock {\em PoS}, ICRC2023:016, 2023.

\bibitem{Jui2011CosmicRI}
Charles Chia~Chung. Jui and for~the Telescope Array~Collaboration.
\newblock Cosmic ray in the northern hemisphere: Results from the telescope array experiment.
\newblock {\em Journal of Physics: Conference Series}, 404:012037, 2011.

\bibitem{PhysRevLett.104.161101}
The High Resolution Fly's Eye~Collaboration et~al.
\newblock Indications of proton-dominated cosmic-ray composition above 1.6 eev.
\newblock {\em Phys. Rev. Lett.}, 104:161101, Apr 2010.

\bibitem{2008ICRC....4..253A}
D.~{Allard}, A.~V. {Olinto}, and E.~{Parizot}.
\newblock {Signatures of the extragalactic cosmic-ray source composition from spectrum and shower depth measurements}.
\newblock In {\em International Cosmic Ray Conference}, volume~4 of {\em International Cosmic Ray Conference}, pages 253--256, January 2008.

\bibitem{2014JCAP...10..020A}
R.~{Aloisio}, V.~{Berezinsky}, and P.~{Blasi}.
\newblock {Ultra high energy cosmic rays: implications of Auger data for source spectra and chemical composition}.
\newblock {\em JCAP}, 2014(10):020--020, October 2014.

\bibitem{aab2017combined}
Alexander Aab, P~Abreu, MARCO Aglietta, Imen Al~Samarai, IFM Albuquerque, Ingomar Allekotte, A~Almela, J~Alvarez Castillo, J~Alvarez-Mu{\~n}iz, GA~Anastasi, et~al.
\newblock Combined fit of spectrum and composition data as measured by the pierre auger observatory.
\newblock {\em Journal of Cosmology and Astroparticle Physics}, 2017(04):038, 2017.

\bibitem{mollerach2019ultrahigh}
Silvia Mollerach and Esteban Roulet.
\newblock Ultrahigh energy cosmic rays from a nearby extragalactic source in the diffusive regime.
\newblock {\em Physical Review D}, 99(10):103010, 2019.

\bibitem{1988A&A...199....1B}
V.~S. {Berezinskii} and S.~I. {Grigor'eva}.
\newblock {A bump in the ultra-high energy cosmic ray spectrum}.
\newblock {\em AAP}, 199(1-2):1--12, June 1988.

\bibitem{BEREZINSKY2014120}
V.~Berezinsky.
\newblock Extragalactic cosmic rays and their signatures.
\newblock {\em Astroparticle Physics}, 53:120--129, 2014.
\newblock Centenary of cosmic ray discovery.

\bibitem{2005PhLB..612..147B}
V.~{Berezinsky}, A.~Z. {Gazizov}, and S.~I. {Grigorieva}.
\newblock {Dip in UHECR spectrum as signature of proton interaction with CMB [rapid communication]}.
\newblock {\em Physics Letters B}, 612(3-4):147--153, April 2005.

\bibitem{2006PhRvD..74d3005B}
Veniamin {Berezinsky}, Askhat {Gazizov}, and Svetlana {Grigorieva}.
\newblock {On astrophysical solution to ultrahigh energy cosmic rays}.
\newblock {\em \prd}, 74(4):043005, August 2006.

\bibitem{harari2014anisotropies}
Diego Harari, Silvia Mollerach, and Esteban Roulet.
\newblock Anisotropies of ultrahigh energy cosmic rays diffusing from extragalactic sources.
\newblock {\em Physical Review D}, 89(12):123001, 2014.

\bibitem{ALOISIO2012129}
R.~Aloisio, V.~Berezinsky, and A.~Gazizov.
\newblock Transition from galactic to extragalactic cosmic rays.
\newblock {\em Astroparticle Physics}, 39-40:129--143, 2012.
\newblock Cosmic Rays Topical Issue.

\bibitem{ALOISIO201373}
R.~Aloisio, V.~Berezinsky, and S.~Grigorieva.
\newblock Analytic calculations of the spectra of ultra-high energy cosmic ray nuclei. i. the case of cmb radiation.
\newblock {\em Astroparticle Physics}, 41:73--93, 2013.

\bibitem{2018JCAP...07..051G}
G.~{Giacinti}, M.~{Kachelrieẞ}, and D.~V. {Semikoz}.
\newblock {Reconciling cosmic ray diffusion with Galactic magnetic field models}.
\newblock {\em JCAP}, 2018(7):051, July 2018.

\bibitem{2019JPhCS1181a2039K}
M.~{Kachelrie{\ss}}.
\newblock {Anisotropic diffusion and the cosmic ray anisotropy}.
\newblock In {\em Journal of Physics Conference Series}, volume 1181 of {\em Journal of Physics Conference Series}, page 012039, February 2019.

\bibitem{kolmogorov1941local}
Andrey~Nikolaevich Kolmogorov.
\newblock The local structure of turbulence in incompressible viscous fluid for very large reynolds.
\newblock {\em Numbers. In Dokl. Akad. Nauk SSSR}, 30:301, 1941.

\bibitem{kraichnan1965inertial}
Robert~H Kraichnan.
\newblock Inertial-range spectrum of hydromagnetic turbulence.
\newblock {\em The Physics of Fluids}, 8(7):1385--1387, 1965.

\bibitem{Jones_2001}
Frank~C. Jones, Andrew Lukasiak, Vladimir Ptuskin, and William Webber.
\newblock The modified weighted slab technique: Models and results.
\newblock {\em The Astrophysical Journal}, 547(1):264, jan 2001.

\bibitem{Roberto_Aloisio_2013}
Roberto Aloisio and Pasquale Blasi.
\newblock Propagation of galactic cosmic rays in the presence of self-generated turbulence.
\newblock {\em Journal of Cosmology and Astroparticle Physics}, 2013(07):001, jul 2013.

\bibitem{Evoli_2018}
Carmelo Evoli, Daniele Gaggero, Andrea Vittino, Mattia~Di Mauro, Dario Grasso, and Mario~Nicola Mazziotta.
\newblock Cosmic-ray propagation with dragon2: Ii. nuclear interactions with the interstellar gas.
\newblock {\em Journal of Cosmology and Astroparticle Physics}, 2018(07):006, jul 2018.

\bibitem{cowsik2016spectral}
Ramanath Cowsik and Tsitsi Madziwa-Nussinov.
\newblock Spectral intensities of antiprotons and the nested leaky-box model for cosmic rays in the galaxy.
\newblock {\em The Astrophysical Journal}, 827(2):119, 2016.

\bibitem{Foreman-Mackey_2013}
Daniel Foreman-Mackey, David~W. Hogg, Dustin Lang, and Jonathan Goodman.
\newblock emcee: The mcmc hammer.
\newblock {\em Publications of the Astronomical Society of the Pacific}, 125(925):306, feb 2013.

\bibitem{reichherzer2022regimes}
Patrick Reichherzer, L~Merten, J~D{\"o}rner, J~Becker~Tjus, MJ~Pueschel, and EG~Zweibel.
\newblock Regimes of cosmic-ray diffusion in galactic turbulence.
\newblock {\em SN Applied Sciences}, 4(1):15, 2022.

\bibitem{Reichherzer_2021}
P.~Reichherzer, L.~Merten, J.~Dörner, J.~Becker Tjus, M.~J. Pueschel, and E.~G. Zweibel.
\newblock Regimes of cosmic-ray diffusion in galactic turbulence.
\newblock {\em {SN} Applied Sciences}, 4(1), dec 2021.

\bibitem{zank1998radial}
GP~Zank, WH~Matthaeus, JW~Bieber, and H~Moraal.
\newblock The radial and latitudinal dependence of the cosmic ray diffusion tensor in the heliosphere.
\newblock {\em Journal of Geophysical Research: Space Physics}, 103(A2):2085--2097, 1998.

\bibitem{wang2022turbulent}
B-B Wang, GP~Zank, L-L Zhao, and L~Adhikari.
\newblock Turbulent cosmic ray--mediated shocks in the hot ionized interstellar medium.
\newblock {\em The Astrophysical Journal}, 932(1):65, 2022.

\bibitem{ackermann2012fermi}
Markus Ackermann, Marco Ajello, WB~Atwood, Luca Baldini, Jean Ballet, Guido Barbiellini, D~Bastieri, K~Bechtol, R~Bellazzini, B~Berenji, et~al.
\newblock Fermi-lat observations of the diffuse $\gamma$-ray emission: implications for cosmic rays and the interstellar medium.
\newblock {\em The Astrophysical Journal}, 750(1):3, 2012.

\bibitem{feng2016bayesian}
Jie Feng, Nicola Tomassetti, and Alberto Oliva.
\newblock Bayesian analysis of spatial-dependent cosmic-ray propagation: Astrophysical background of antiprotons and positrons.
\newblock {\em Physical Review D}, 94(12):123007, 2016.

\bibitem{Yang:2022jck}
H.~Y.~Karen Yang, Mateusz Ruszkowski, and Ellen~G. Zweibel.
\newblock {Fermi and eROSITA bubbles as relics of the past activity of the Galaxy\textquoteright{}s central black hole}.
\newblock {\em Nature Astron.}, 6(5):584--591, 2022.

\bibitem{cao2023measurement}
Zhen Cao, F~Aharonian, Q~An, YX~Bai, YW~Bao, D~Bastieri, XJ~Bi, YJ~Bi, JT~Cai, Q~Cao, et~al.
\newblock Measurement of ultra-high-energy diffuse gamma-ray emission of the galactic plane from 10 tev to 1 pev with lhaaso-km2a.
\newblock {\em Physical Review Letters}, 131(15):151001, 2023.

\bibitem{PhysRevD.71.083007}
Martin Lemoine.
\newblock Extragalactic magnetic fields and the second knee in the cosmic-ray spectrum.
\newblock {\em Phys. Rev. D}, 71:083007, Apr 2005.

\bibitem{PhysRevD.104.063005}
Juan~Manuel Gonz\'alez, Silvia Mollerach, and Esteban Roulet.
\newblock Magnetic diffusion and interaction effects on ultrahigh energy cosmic rays: Protons and nuclei.
\newblock {\em Phys. Rev. D}, 104:063005, Sep 2021.

\bibitem{2023ApJ...957...80C}
Antonio {Condorelli}, Jonathan {Biteau}, and Remi {Adam}.
\newblock {Impact of Galaxy Clusters on the Propagation of Ultrahigh-energy Cosmic Rays}.
\newblock {\em \apj}, 957(2):80, November 2023.

\bibitem{10.1111/j.1365-2966.2008.14132.x}
J.~Donnert, K.~Dolag, H.~Lesch, and E.~Müller.
\newblock {Cluster magnetic fields from galactic outflows}.
\newblock {\em Monthly Notices of the Royal Astronomical Society}, 392(3):1008--1021, 01 2009.

\bibitem{murase2012blazars}
Kohta Murase, Charles~D Dermer, Hajime Takami, and Giulia Migliori.
\newblock Blazars as ultra-high-energy cosmic-ray sources: implications for tev gamma-ray observations.
\newblock {\em The Astrophysical Journal}, 749(1):63, 2012.

\bibitem{bai2022large}
X~Bai, BY~Bi, XJ~Bi, Z~Cao, SZ~Chen, Y~Chen, A~Chiavassa, XH~Cui, ZG~Dai, D~Della~Volpe, et~al.
\newblock The large high altitude air shower observatory (lhaaso) science book (2021 edition).
\newblock {\em Chin. Phys. C}, 46:035001--035007, 2022.

\bibitem{PhysRevD.99.103010}
Silvia Mollerach and Esteban Roulet.
\newblock Ultrahigh energy cosmic rays from a nearby extragalactic source in the diffusive regime.
\newblock {\em Phys. Rev. D}, 99:103010, May 2019.

\bibitem{10.1143/PTP.113.721}
Hiroshi Muraishi, Shohei Yanagita, and Tatsuo Yoshida.
\newblock {Galactic Modulation of Extragalactic Cosmic Rays: — A Possible Origin of the Knee in the Cosmic Ray Spectrumc —}.
\newblock {\em Progress of Theoretical Physics}, 113(4):721--731, 04 2005.

\bibitem{osti_6709451}
S~P Ahlen, P~B Price, M~H Salamon, and G~Tarle.
\newblock Can we detect antimatter from other galaxies.
\newblock {\em Astrophys. J.; (United States)}, 260:1, 9 1982.

\bibitem{Subedi_2017}
P.~Subedi, W.~Sonsrettee, P.~Blasi, D.~Ruffolo, W.~H. Matthaeus, D.~Montgomery, P.~Chuychai, P.~Dmitruk, M.~Wan, T.~N. Parashar, and R.~Chhiber.
\newblock Charged particle diffusion in isotropic random magnetic fields.
\newblock {\em The Astrophysical Journal}, 837(2):140, mar 2017.

\bibitem{taylor1922diffusion}
Geoffrey~I Taylor.
\newblock Diffusion by continuous movements.
\newblock {\em Proceedings of the london mathematical society}, 2(1):196--212, 1922.

\bibitem{Kubo1957}
Kubo R.
\newblock Statistical-mechanical theory of irreversible processes. i. general theory and simple applications to magnetic and conduction problems.
\newblock {\em Journal of the physical society of Japan}, 12(6):570--586, 1957.

\end{thebibliography}
\newpage
\end{document}